\documentclass[runningheads]{llncs}

\usepackage{amssymb}
\usepackage{graphicx}
\begin{document}
\mainmatter  
\title{Quantum amplitude amplification algorithm: an explanation of availability bias}
\titlerunning{Quantum amplitude amplification algorithm and availability bias}

%
%
\author{Riccardo Franco}
\authorrunning{Quantum amplitude amplification algorithm and availability bias}

\institute{ }
%
%

\toctitle{Lecture Notes in Computer Science}
\tocauthor{Authors' Instructions}
\maketitle
\begin{abstract}
In this article, I show that a recent family of quantum algorithms, based on the quantum amplitude amplification algorithm, can be used to describe a cognitive heuristic called availability bias. The amplitude amplification algorithm is used to define quantitatively the ease of a memory task, while the quantum amplitude estimation and the quantum counting algorithms to describe cognitive tasks such as estimating probability or approximate counting. 
\end{abstract}

%

\section{Introduction}
The idea that human judgements and decision-making can evidence
quantum mechanics behavior has a great deal of intuitive appeal, and
it is at the basis of a recent research topic, which can be called
quantum cognition. A number of authors have explored such idea, like
\cite{Busemeyer} for decision making, or \cite{Rfranco_conj2} and
\cite{inverse_fallacy_franco} for human judgements. The quantum-like
models there proposed seem to adequately describe the experimental
results: however, the potentialities of the quantum formalism have
not fully explored, mainly for what concerns the quantum parallelism
and a characterization of quantum algorithms in terms of human
tasks.

In the present article, I propose to describe the experimental results concerning the availability bias with
the \textit{quantum amplitude amplification}, \textit{quantum amplitude
estimation} and \textit{quantum counting }algorithms \cite{Counting}: the first is a recent
generalization of the Grover's algorithm \cite{Grover1}, while the other two algorithms are
applications of the amplitude amplification, followed by a quantum
Fourier transform. I show that these algorithms are able to model
some important experimental results of cognitive science relevant to availability bias: in
particular, the amplitude amplification algorithm allows to give a mathematical characterization of the
ease to recall items or concepts, while the
amplitude estimation/counting algorithms allow to introduce a formal connection between such ease and judgements of
probability/frequency about facts.
Here I do not discuss about the physical possibility for human mind
to perform quantum algorithms: I only consider from a formal point
of view the problem of computational complexity and the possibility
to define mathematicaly the ease to remember. The
Grover's algorithm  is an important quantum algorithm based on
parallelism which allows to search in an unsorted database with a
high number of items faster than any classical algorithm (quadratic
speedup). One of the first attempts to use such algorithm in
cognitive science (more precisely a generalization \cite{Grover2})
has been done by Franco \cite{Rfranco_memory} to describe the
influence of emotions on the ease to remember.
The same quadratic speedup is provided by the
algorithms based on the amplitude amplification algorithm,
based on quantum parallelism. 

This article attempts to model within the quantum framework the
\textit{availability bias}, a human cognitive bias that causes
people to  estimate  frequency or probability on the basis of
\textit{how easily} they can recall or imagine instances of whatever
it is they are trying to estimate. The availability bias, which is
at the root of many other human biases and culture-level effects,
was discovered  by psychologists Amos Tversky and Daniel Kahneman
(2002 Nobel Prize in Economics) \cite{Kahn-Tversky-availability}.
A simple example of the availability bias, which I discuss in the present article,  is provided
by a famous experiment of Tversky and Kahneman (1973) \cite{Kahn-Tversky-availability}:
\textit{Consider the letter R: is R more likely to appear in the
first position or in the third position?}
The most part of participants judged the first
position to be more likely. However in the English language there are more
words with R at the third position than at the first position. The
explanation given by Tversky and Kahnemann is that people estimate
the number of words based on the ease with which they can recall
them, which is the availability heuristic:
the first letter provides a better cue for recalling instances of
words than does the third letter.
It is evident from the latter example that the judgements performed
by people about the words involves in theory a great number of
computations. In fact, the English language contains about 500000
words, and the task previously described involves in theory computations over
such set.
This  considerations make stronger the quantum-like point of view,
since the quantum algorithms here proposed manifest a quadratic
speed up, and thus are faster than any classic algorithm.


\section{Availability bias}
Availability bias is a human cognitive bias that causes people to estimate
the probability or the number of particular categories of
items on the basis of \textit{how easily} they can recall or imagine them.
In the definition of the availability bias is important  to
operationalize the ease with  which the  memory processes are
performed. In particular, two different definitions have been used
widely in availability experiments,  giving an experimental measure
of the ease of the memory task: 1) \textit{availability-by-number}: the produced proportion of good items versus bad items in a fixed time; 2) \textit{availability-by-speed}:  the time ratio between the consumed
retrieval times for the same number of good items and bad items.
In general, the availability  experiments involve two
different groups of subjects: one which performs the memory 
task, and one that performs the judgements about the
probability/number of items. Thus the availability experiments
 verify a positive correlation between the measure of ease
in the memory task (by using the availability-by-number or the
availability-by-speed) and the quantitative judgements performed by
the subjects.
I will focus the attention on  the following two categories of
experiments: 
\\
1) Judgements of probability: the availability effects can be due to
the ease of recalling items (as in the example of
\cite{Kahn-Tversky-availability} described in the introduction about
the likelihood of letter $R$ in the first or in the third position
of English words) or to the vividness of particular events in
memory: for example, in \cite{combs} subjects have been asked to
estimate the probability of plane accidents (quite events
rare, even if the vast majority of the population overestimates
their probability).
\\
2) Judgements of  number: in \cite{Kahn-Tversky-availability} is
described a simple recalling experiment, where subjects were
presented a recorded list of names of known personalities of both
sexes. After listening to the list, some subjects judged whether it
contained more names of men or women, others attempted to recall the
names in the list. In particular, the list included 19 names of
famous personalities of one sex and 20 names of less famous
personalities of the other sex. The experimental results show a
positive correlation between the estimated number of persons of the
more famous group and the number of recalled names of the same
group.
\section{Amplitude amplification algorithm}
The amplitude amplification algorithm \cite{Counting} is a generalization of
Grover's algorithm, and it can be used for solving the following problem: let us
consider $N$ items and a boolean function $f:\{0, 1, ... ,
N-1\}\rightarrow \{0, 1\}$, which partitions the items into $t$ \textit{good items}
 (those for which $f$ is equal to 1), and $N-t$ \textit{bad items}
(those for which $f$ is equal to 0).
It  is evident that such algorithm can be used to model
the retrieval tasks in cognitive science. For example, the
experiment of  Tversky and Kahneman \cite{Kahn-Tversky-availability}
relevant to words with letter $R$ in first or third position can be
represented as a partitioning of English words in two categories:
the good items (words with $R$ at first position) and the bad items
(words with $R$ at third position).
Even if the mathematical details of the algorithm are described in
next subsection, I now present the main features, reducing to the minimum the
formalism. The intuitions here presented are similar to those
preliminarly exposed in Franco \cite{Rfranco_memory}.
The quantum amplification algorithm, like the Grover's algorithm, is
composed by three main parts:
\\
1) The \textit{initial state}, in which the $N$ items 
are encoded into the elements of a basis of a
$N$-dimensional vector space. An important feature of the amplitude
amplification algorithm, which differences it from the Grover's
algorithm, is that the items within such initial state can have
different weights: in particular, the parameter $a$ is the
probability to measure a good item in such initial state. In
Grover's algorithm we always have $a=t/N$. The initial state can be
interpeted, in the context of cognitive processes, as a
\textit{guessing state}, representing the initial mental weights
relevant to the items. If $a>t/N$, this means that the good items
have initially more relevance than the bad items. If the guessing
state is a flat distribution over all the items ($a=t/N$), this means that the subjects have no preliminar idea about good/bad items.
\\
2) The \textit{amplification engine}, which is an  iterative process
allowing to enhance the weights of the good items: at each step the boolean function $f$ is
evaluated simultaneously over all the items, and the weights of the
good items are enhanced through interference effects. Differently
from the Grover's algorithm, the efficiency of the amplification
engine depends on the
\textit{guessing state}: the algorithm succeeds
after a number of iterations  proportional to $1/\sqrt{a}$. If
$a=t/N$ the algorithm is equal to the Grover's algorithm, and the
required number of steps is proportional to $\sqrt{N/t}$. It is
important to note that a classic algorithm would imply a number of
steps proportional to $N/t$, while the Grover's algorithm allows for
a quadratic speedup, that is a number of steps proportional to
$\sqrt{N/t}$. The amplitude amplification algorithm allows for a
further speedup when the guessing state is such that $a>t/N$,
because the number of required steps is proportional to
$1/\sqrt{a}<\sqrt{N/t}$: the initial guessing state
gives higher weight to the good items than to the bad items, making faster the retrieval process.
\\
The interpretation  of such amplification engine in the context of
cognitive tasks is in terms of subconscious processes: they allow
for parallelism in the evaluation of the boolean function over all
the items, but they need a number of iterations  proportional to
$1/\sqrt{a}$ to amplify the probability of good items. In other
words, the subjects are able to apply $f(x)$ on each item $x$ (thus
deciding if each item is good or bad). The algorithm suggests that such decision procedure
is performed in a parallel and subconscious way, thus faster
than in a serial way. 
\\
3) A \textit{measure} on the final state. The algorithm modifies the initial guess state,
producing a final state which contains almost only good states. Thus
a final measure produces one of the good items, and the
recall task is finished. This fact represents in my description the
conscious act of remembering.
\\
The amplitude  amplification algorithm allows to give a simple
mathematical definition of the \textit{ease} to retrieve in terms of
the {availability-by-speed}: the time required to find a good item
is proportional to $1/\sqrt{a}$, where $a$ is the initial guessing
parameter: a high value of $a$ gives a short time to retrieve a
good item. The parameter $a$ represents
how vivid are the good items in memory before retrieving them: it
can change depending on attempts of imagining instances
of the retrieved items. Analogously, the {availability-by-number} is
the number of good items that subjects can remember in a fixed time:
it is proportional, in our model, to $\sqrt{a}$.
In the experiment on the position of letter $R$ in English words \cite{Kahn-Tversky-availability},
the time to produce the word is lower with $R$ as first letter than as the third letter. Thus
I assume that the guess state contains a set of $N$ items (the most
common English words), and the weight for the words beginning with
$R$ is higher than for those with $R$ at third position.
%
%
\subsection{Mathematical details for the amplitude amplification algorithm}\label{m1}
In the quantum formalism, the partition of $N$ items into good and bad items
leads to consider a $N-$dimensional Hilbert space, whose
computational basis is $\{|0\rangle, |1\rangle, ..., |N-1\rangle\}$:
each vector corresponds to a particular item. Thus the function $f$
introduces a partition of $H$ into a \textit{good subspace} (spanned
by the vectors $|x\rangle$ for which $f(x)=1$) and a \textit{bad
subspace} (spanned by the vectors $|x\rangle$ for which $f(x)=0$).
Thus any superposition $|s\rangle =\sum_x \psi(x)|x\rangle$ can be written as
$|s\rangle=|\psi_0\rangle + |\psi_1\rangle$, where $|\psi_1\rangle$
is the superposition of good vectors ($f(x)=1$) and $|\psi_0\rangle$ is the superposition of bad vectors ($f(x)=1$).\\
The algorithm presents the following steps: \\
1) \textit{Initial state}: prepare the vector
$A|0\rangle=|\psi_0\rangle + |\psi_1\rangle$, where $A$ is a quantum
algorithm which uses no measurement, and $a=\langle
\psi_1|\psi_1\rangle$ is the probability to measure a good state. If
$A$ is the quantum Fourier transform $F_N: |x\rangle  \rightarrow
N^{-1/2} \sum_{y=0}^{N-1}e^{2\pi i x y} |y\rangle$, we have a
uniform superposition of vector states with amplitude $N^{-1/2}$,
and $a=t/N$ (as in standard Grover's algorithm).
\\
2) \textit{Amplification engine}: apply the operator $Q=-A S_{0}
A^{-1}S_{f}$, where $S_0$ and $S_{f}$ are conditional phase
inversion  operators ($S_0$ changes the sign of the amplitude if and
only if the state is the zero state $|0\rangle$, while $S_{f}$
conditionally changes the sign of the amplitudes of the good
states).
\\
3) \textit{Measure}  the final state: obtain one of the search
results, measuring the resulting state in the computational basis.
\\
It can be shown that after
$
\left\lfloor
\pi/4 arcsin(\sqrt{a})
\right\rfloor
$ iterations
(where $\left\lfloor x \right\rfloor$ is the rounding of $x$) the
measured outcome is good with probability at least $max(a,1-a)$. If
we have a high number of items $N$ and $a\ll N$, then the optimal
number of iterations is proportional to $1/\sqrt{a}$. If $A$ is the
quantum Fourier transform the optimal number of iterations is
proportional to $\sqrt{N/t}$, which corresponds to the speedup of
Grover's algorithm. If  $a>t/N$, the number of iterations is
lower than $\sqrt{N/t}$. 
%
\section{The quantum amplitude estimation algorithm}
The quantum amplitude estimation algorithm  \cite{Counting} allows
to estimate the amplitude of a quantum state by applying at
different steps the amplitude amplification algorithm. 
From a cognitive point of view, it allows to estimate
with a good precision the probability $a$ to find a good item 
(according to the partitioning introduced by finction $f$)
when the opinion state about the $N$ items is the initial guessing state.
Even if the mathematical  details of the algorithm are described in next subsection, I now present its main features, reducing to the minimum the formalism. The algorithm can be decomposed in three parts:\\
1) \textit{Initial state}: it is composed by the guessing state, as described before.
\\
2) \textit{Parallel amplifications}:  different instances of the
amplification engine are applied in a parallel way, with different
numbers of iterations. Thus we have a double level of parallelism:
in each step of the amplification engine the function $f(x)$ is
applied simultaneously to the items, and this works simultaneously
for each instance of the amplification engine.
\\
3) \textit{Analysis} of the different  amplifications: since the
efficiency of each amplification engine depends on the parameter
$a$, the analysis of different instances of the amplification
process with different number of iterations allow to estimate $a$,
with a few standard deviations, after a number of evaluations of $f$
proportional to $1/\sqrt{a}$.
\\
This algorithm is particularly important  for the study of cognitive
processes, because it allows to describe the tasks where subjects
produce subjective probabilities relevant to events. In this sense,
it provides the formal link between a quantum-like approach
describing choices (for example, \cite{Busemeyer}) and a
quantum-like approach describing subjective probabilities (for
example, \cite{Rfranco_conj2}): choices are the effect of simple
measurements on quantum states, while the subjective probabilities
are the result of a quantum amplitude estimation algorithm applied
on the same state.
In the context of availability bias, the present algorithm  can be
used to describe the experiment of \cite{Kahn-Tversky-availability}
presented in the introduction about the likelihood of letter $R$ in
the first or in the third position of English words. The retrieve
process for words with $R$ in first or third position involves two
different partitioning of English words and two different
amplification processes with parameters $a$ and $a'$.  In other
words, we assume that subjects' mental state (the guess state)
involves $N$ words, and that the weight in such state relevant to
words with $R$ in first and third position is $a$ and $a'$
respectively. According to our model, the ease to recall words with
$R$ in first position can be described by the availability-by-number
and is proportional to $\sqrt{a}$, and the estimated probability to
recall words with $R$ in first position is near to $a$. Thus if
subjects recall more words with $R$ in first position than in third
($\sqrt{a}>\sqrt{a'}$), then the estimated probability to find a
word with letter $R$ in first position is higher than the estimated
probability to find words with $R$ in third position ($a>a'$).
The same formalism can  be used to describe the experiments in
\cite{combs}, where subjects overestimated the probability of plane
accidents, because of  the vividness of such events in memory.
\\
Like for the amplitude  amplification algorithm, also in this case
the produced estimated probability can be described as the result of
subconscious amplification processes (with evaluations of function
$f$) and a final analysis and measure.
%
%
\subsection{Mathematical description of amplitude estimation algorithm}\label{m2}
The amplitude estimation algorithm,  called \textbf{Est\_Amp}($A, f, M$), is able to estimate the amplitude of $|\psi_1\rangle$  (good states superposition) in $A|0\rangle$. It is based on the amplitude amplification algorithm. In particular:\\
\textit{1) Initial state}: prepare the vector $F_{M}|0\rangle
A|0\rangle$,  formed by two distinct registers: the first has
dimension $M$,  while the second has dimension $N$. We recall that
$F_M$ is the quantum Fourier transform $F_M: |x\rangle \rightarrow
M^{-1/2} \sum_{y=0}^{M-1}e^{2\pi i x y} |y\rangle$.
\\
\textit{2) Parallel amplifications}:  apply the operator
$\Lambda_{M}(Q)$, defined by $|j\rangle |y\rangle \rightarrow
|j\rangle Q^{j}|y\rangle$  with $0 \leq j\leq M$,  where $Q=-A S_0
A^{-1} S_f$ is the standard amplitude amplification engine. In other
words, operator $\Lambda_{M}(Q)$ applies in a parallel way different
degrees of amplification, from 0 to $M$, to the guess state
$A|0\rangle$.
\\
\textit{3) Find the period of the wave function}: apply $F^{-1}_{M}$
to the first register and measure it, obtaining  an integer $y$. The
estimated amplitude is then $\tilde{a}=sin^{2}(\pi y/M)$ with a good
approximation: the accuracy of such estimate is given in Theorem 12
in \cite{Counting}. In particular,  to obtain a probability estimate
with a few standard deviations, we have to choose $M=\left\lfloor
1/\sqrt{a} \right\rfloor$.
%
\section{The quantum counting algorithm}
The quantum counting
algorithm \cite{Counting} allows, given a boolean
function $f$ defined on a set $X$ of $N$ items, to estimate the number of
elements of $X$ for which the function $f$ is true
$t=\left| \{x\in X|f(x)=1 \right|$.
In other words, the algorithm allows to estimate the size of the subset of
\textit{good items} (those for which $f(x)=1$).
The best classical strategy is to evaluate $f$ on random elements of
$X$: thus the number of evaluations in order to have a good estimate
of $t$ is proportional to $N$. On the contrary, the quantum counting
algorithm allows to produce good estimates for such number in approximatively $\sqrt{N}$ steps  (quadratic speedup).

The quantum counting algorithm can be  considered as an application
of the previous amplitude estimation algorithm. In fact, if the
guessing state assigns the same weight to all the items, then the
estimated probability relevant to the  good items  is near to $t/N$:
the approximate number of good items can be obtained by multiplying
such estimated probability by $N$. I propose here a simple
generalization of the quantum counting algorithm, which I will
discuss in mathematical details in the next subsection: if the
guessing state assigns non-uniform weights to the items, the
probability relevant to good items is $a\neq t/N$. If for example
$a>t/N$, then the estimated number of items is near to $aN>t$: we
have an overestimation of the number of items, due to the guessing
state in the amplification process.

Such simple generalization allows to describe the recalling
experiment in \cite{Kahn-Tversky-availability} , where subjects were
presented a recorded list of names of known personalities of both
sexes. After listening to the list, some subjects judged whether it
contained more names of men or women, others attempted to recall the
names in the list. In particular, the list included 19 names  of
famous personalities of one sex and 20 names of less famous
personalities of the other sex. The experimental results show a
positive correlation between the estimated number of persons of the
more famous group and the number of recalled names of the same
group. In fact the same parameter $a$ is involved both in the
recalling process and in the approximate counting process: thus the
ease to recall names of one group (proportional to $\sqrt{a}$)
entails a higher estimated size of the same group ($aN$).

%
\subsection{Mathematical description of quantum counting algorithm}\label{m3}
Given a boolean function $f$ over a discrete set $X$ with $N$
elements, the quantum counting algorithm \textbf{Count}($F_N, f,
M$)  can be written as a special case of the amplitude estimation:
$\tilde{t}=N \times \textbf{Est\_Amp}(F_N, f, M)$.
If we  use, instead of the Fourier transform $F_N$, a generic
operator $A$, the quantum counting algorithm \textbf{Count}($A, f,
M$) does not produce a correct estimate of $t$, the number of good
items. However, if $a>t/N$, the modified counting algorithm produces
an estimate $\tilde{t}>t$, while if $a<t/N$, it produces an estimate
$\tilde{t}<t$. 
%
%
\section{Conclusions}
In this article I show how three important quantum algorithms can model the experimental results of 
availability bias. I introduce the amplitude
amplification algorithm to give a mathematical characterization of
the ease to recall items or concepts. Then I present the amplitude
estimation/counting algorithm, establishing a connection
between the ease to retrieve and the judgements of
probability/frequency about facts. 
The quantum description of availability bias, and in particular the
use of quantum algorithms, has some advantages: 1) the economy of a
quantum description, which seems to be consistent with a large
number of cognitive heuristics (see for example \cite{Rfranco_conj2},
and \cite{Busemeyer}), while
the classic alternatives are ad-hoc
models with a very weak mathematical
apparatus; 2) as noted by Manin \cite{Manin}, some human tasks, such
as playing chess or speech generation and perception, require a
great number of computations per second, as is evidenced by
efficient chess playing software (based on classical algorithms).
Since the characteristic time of neuronal processing is about
$10^{-3}$ seconds, it seems difficult that a classical model could
describe such tasks: in the experiments of
words with letter $R$ of \cite{Kahn-Tversky-availability}, the set
of words on which perform the computation is in theory of 500000
elements, thus making a classic algorithm modeling the cognitive
processes more difficult to apply than fast quantum algorithms.
\\
There are some questions which need further investigations:
1) How the amplitude estimation processes can be influenced  by a
change in the partitioning? For example, we can choose to partition
the items into two different ways.
2) Availability bias is relevant not only with estimated
probabilities or approximate counting, but also with generic
evaluations, like described in \cite{fox}. It
should be investigated if the algorithm used in the present
algorithm can be generalized also to generic evaluations (like for
example course ratings).
%
%
\end{document}